\documentclass[preprint,prb,aps,showpacs,epsfig]{revtex4}
\usepackage{amsfonts}
\usepackage{amsmath}
\usepackage{amssymb}
\usepackage{graphicx}

\begin{document}

\title{Analytical spectral-domain scattering theory of a general gyrotropic sphere}


\author{Youlin Geng$^{1}$ and Cheng-Wei Qiu$^{2,*}$}

\affiliation{$^{1}$The Institute of Antenna and Microwaves, Hangzhou Dianzi University, Hangzhou, Zhejiang, China 310018. \\
$^{2}$Department of Electrical and Computer Engineering, National
University of Singapore, 4 Engineering Drive 3, Singapore 117576.
Tel: +6565162259. Fax: +6567791103. *e-mail: eleqc@nus.edu.sg) }
\date{\today}

\bibliographystyle{prsty}


\newcommand{\bq}{\begin{subequations}}
\newcommand{\eq}{\end{subequations}}
\newcommand{\be}{\begin{equation}\displaystyle}
\newcommand{\ee}{\end{equation}}
\newcommand{\beq}{\begin{eqnarray}}
\newcommand{\eeq}{\end{eqnarray}}
\newcommand{\beqq}{\begin{eqnarray*}}
\newcommand{\eeqq}{\end{eqnarray*}}

\newcommand{\mt}{\mbox{$\hspace{-1.5ex}$}}              
\newcommand{\mT}{\mbox{$\hspace{-3.0ex}$}}              
\newcommand{\st}[1]{{\mbox{${\mbox{\scriptsize #1}}$}}}  
\newcommand{\BM}[1]{\mbox{\boldmath $#1$}}              
\newcommand{\uv}[1]{\mbox{$\widehat{\mbox{\boldmath $#1$}}$}}   
\newcommand{\dy}[1]{\mbox{\boldmath $\overline{#1}$}}   
\newcommand{\sbtex}[1]{{\mbox{\scriptsize #1}}}   
\newcommand{\nab}{\mbox{\boldmath $\nabla$}}            
\newcommand{\CS}{\mbox{$\begin{array}{c}\cos \\ \sin \\ \end{array}$}}
\newcommand{\SC}{\mbox{$\begin{array}{c}\sin \\ \cos \\ \end{array}$}}
\def\centertiff#1#2#3{\vskip#2\relax\centerline{\hbox to#1{\special
  {bmp:#3 x=#1, y=#2}\hfil}}}
\def\twoeps#1#2#3#4{\vskip#2\relax\centerline{{\hbox to#1{\special
  {eps:#3 x=#1, y=#2}\hfil}}\hspace{-1ex}
  {\hbox to#1{\special {eps:#4 x=#1, y=#2}\hfil}}}}

\newcommand{\citeasnoun}[1]{Ref.~\citenum{#1}}

\def\a{s}
\def\b{s}
\newcommand{\add}[1]{\if\a\b{{\color{red} #1}}\else{#1}\fi}
\newcommand{\comm}[1]{\if\a\b{\marginpar{\color{blue}\{\small #1\}}}\else{}\fi}
\newcommand{\del}[1]{{\if\a\b{{\color{magenta}[[#1]]}}\else{}\fi}}





\begin{abstract}
We propose an analytical scattering theory in spectral domain to
model the electromagnetic (EM) fields of a gyrotropic sphere in
terms of the eigen-functions and their associated spectral
eigenvalues/coefficients in a recursive integral form. Applying the
continuous boundary conditions of electromagnetic fields on the
surface between the free space and gyrotropic sphere, the spectral
coefficients of transmitted fields inside the gyrotropic sphere and
the scattered fields in the isotropic host medium can be obtained
exactly by expanding spherical vector wave eigenfunctions. Numerical
results are provided for some representative cases, which are
compared to the results from adaptive integral method (AIM). Good
agreement demonstrates the validity of the proposed analytical
scattering theory for gyrotropic spheres in spectral domain using
Fourier transform.
\end{abstract}

\maketitle
\newpage

\section{Introduction}
Electromagnetic scattering of anisotropic media have attracted more
and more attention for their wide applications in the past decades,
such as radar cross section (RCS) computation of perfect electric
conductor (PEC) targets coated with complex material, radome design,
and interaction of light/wave with biological media and
metamaterials
\cite{ref1,ref2,ref3,ref6,ref7,ref8,ref9,ref9a,ref9b,ref11,ref12}.

Based on the plane wave expansion in terms of spherical vector wave
functions in isotropic medium \cite{ref13}, the scattering by a
uniaxial sphere and a sphere of uniaxial left-handed materials have
been derived \cite{ref11,ref12}. More recently, the scattering of a
gyromagnetic sphere has been investigated in the expansion in
spatial domain \cite{ref12b}. Moreover, the theory is only working
for the case having gyrotropic permeability and scalar permittivity.
If both permittivity and permeability are gyrotropic matrices, the
interplay between the extra three parameters in the gyrotropic
permittivity will make that approach \cite{ref12b} too tedious and
insufficient to model the scattering properties. This motivates our
work in spectral domain instead of spatial domain. A most general
gyrotropic sphere is considered, and since the existing method has
drawbacks in the analysis of scatterings, a novel approach has to be
developed. The analytical method, which can be readily implemented
by programming, has its academic and practical significance in
contrast to purely numerical solutions from FDTD, FEM or others.

In view of this, we propose a distinguished method based on Fourier
transform, and thus the spectral-domain analysis of the scattering
by a general gyrotropic sphere in terms of spherical functions wave
functions is investigated. This method has distinguished features:
(1) it can straightforwardly be employed to describe the light wave
interaction with particles and objects with gyrotropic permittivity
and permeability; (2) the material constitution is very complex and
general (both $\dy{\epsilon}$ and $\dy{\mu}$ are gyrotropic
tensors), so all those existing scattering theorems are just its
sub-cases, e.g., uniaxial, plasma, anisotropic, gyromagnetic, etc.;
(3) it directly solves for the eigen-problems in spectral domain by
Fourier transform, which simplifies the formulation in spatial
domain \cite{ref12b}.

To obtain the solution of vector wave functions in gyrotropic
anisotropic media, we start from the vector wave equation in a
source-free gyrotropic anisotropic medium. Taking the Fourier
transform of the electric field and substituting it into the vector
wave equation of the electric field, we obtain the characteristic
equation. Solving this equation, the eigenvalues and corresponding
vector wave eigenfunctions can be yielded. Then, electromagnetic
fields inside and outside the gyrotropic anisotropic sphere can be
expressed based on the eigenvalues and eigenfunctions. Those unknown
scattering coefficients can be analytically determined from applying
the continuous boundary conditions on the surface of the gyrotropic
anisotropic sphere, where orthogonality relations of the Legendre
polynomials are employed. Numerical results are obtained to gain
more physical insight into this problem. After the results were
validated by comparison with the existing data, some new results are
computed and discussed.

In the subsequent analysis, a time dependence of the form
exp$(-i\omega t)$ is assumed for the electromagnetic field
quantities but is suppressed throughout the treatment.

\section{Analytical formulation}
\begin{figure}[htbh]
\centering\includegraphics[width=9cm]{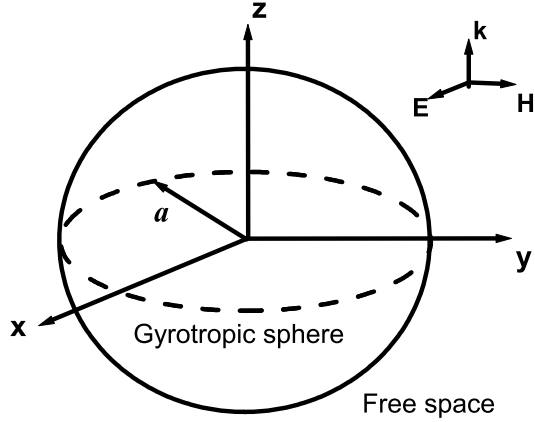} \caption{Geometry
for the EM scattering of a plane wave by an gyrotropic anisotropic
sphere.} \label{fig1}
\end{figure}

The permittivity and permeability tensors of the gyrotropic
anisotropic sphere shown in Fig.~\ref{fig1} are characterized by the
following two matrices \beq \dy{\epsilon} &=&
 \left[ \begin{array}{ccc} \epsilon_{1} & -i\epsilon_{2} & 0 \\  i\epsilon_{2} & \epsilon_{1} & 0 \\ 0 & 0 &\epsilon_{3}\end{array}\right] \nonumber \\
\dy{\mu} &=&
  \left[
\begin{array}{ccc} \mu_{1} & -i\mu_{2} & 0 \\  i\mu_{2} & \mu_{1} & 0 \\ 0 & 0
&\mu_{3}\end{array}\right]. \eeq

The parameters are defined in Cartesian coordinates. The
$\BM{E}$-field vector wave equation can be obtained by substituting
the above constitutive relations into the source-free Maxwell's
equations
\cite{ref12}, i.e., 
\begin{equation}\label{E_quation}
\nab \times \left[ \dy{\mu}^{-1}\cdot \nab \times \BM{E}(\BM{r})
\right] -\omega^{2}\dy{\epsilon}\cdot \BM{E}(\BM{r})=0.
\end{equation}
The solution to (\ref{E_quation}) can be obtained by the following
Fourier transform:
\begin{equation}
\BM{E}(\BM{r})=\int^{\infty}_{-\infty}dk_{x}\int^{\infty}_{-\infty}dk_{y}
\int^{\infty}_{-\infty} \BM{E}(\BM{k}) e^{i\BM{k} \cdot \BM{r}}
dk_{z} \label{fourier}
\end{equation}
where the wave number is denoted by
$\BM{k}=k_{x}\uv{x}+k_{y}\uv{y}+k_{z}\uv{z}$, and the space vector
is identified as  $\BM{r} =x\uv{x} +y\uv{y} +z\uv{z}$, with
$\uv{x}$, $\uv{y}$, $\uv{z}$ being the unit vectors in Cartesian
coordinates. By substituting (3) into (2), the wave equation can be
transformed into
\begin{equation}
\int^{\infty}_{-\infty}dk_{x}\int^{\infty}_{-\infty}dk_{y}
\int^{\infty}_{-\infty}\dy{K}(\BM{k})\cdot \BM{E}(\BM{k}) e^{i\BM{k}
\cdot \BM{r}} dk_{z} = 0 \label{integral}
\end{equation}
where
\begin{equation}
\dy{K}(\BM{k})= \\
\left[\begin{array}{ccc}
-b_{1}k_{z}^{2}-b_{3}k^{2}_{y}+a_{1} & b_{3}k_{x}k_{y}-ib_{2}k^{2}_{z}-ia_{2} & b_{1}k_{x}k_{z}+ib_{2}k_{y}k_{z} \\
b_{3}k_{x}k_{y}+ib_{2}k^{2}_{z}+ia_{2} & -b_{1}k_{z}^{2}-b_{3}k^{2}_{x}+a_{1} & b_{1}k_{y}k_{z}-ib_{2}k_{x}k_{z} \\
b_{1}k_{x}k_{z}-ib_{2}k_{y}k_{z} & b_{1}k_{y}k_{z}+ib_{2}k_{x}k_{z}
& -b_{1}(k_{y}^{2}+k^{2}_{x})+a_{3}
\end{array} \!\! \right]
\label{matrix}
\end{equation}
with \beq
a_{1} &=& \omega^{2}\epsilon_{1}, \nonumber \\
a_{2} &=& \omega^{2}\epsilon_{2}, \nonumber \\
a_{3} &=& \omega^{2}\epsilon_{3}, \nonumber \\
b_{1} &=& \displaystyle\frac{\mu_{1}}{\mu^{2}_{1}-\mu^{2}_{2}}\nonumber \\
b_{2} &=& \displaystyle\frac{\mu_{2}}{\mu^{2}_{1}-\mu^{2}_{2}}\nonumber \\
b_{3} &=& 1/\mu_{3}. \eeq In order to get nontrivial solutions of
$\BM{E}(\BM{k})$, the following characteristic equation has to be
satisfied:
\begin{equation}
\mbox{Det} \left[ \dy{K}(\BM{k}) \right] =0.
\end{equation}
It can be explicitly rewritten as
\begin{equation}\label{biquadratic}
A(\theta_{k},\phi_{k})k^{4}-B(\theta_{k},\phi_{k})k^{2}+C=0,
\end{equation}
where \beq
A(\theta_{k},\phi_{k}) &=& \left[b_{1}b_{3}\sin^{2}\theta_{k}+(b^{2}_{1}-b^{2}_{2})cos^{2}\theta_{k}\right] \times\left[a_{1}\sin^{2}\theta_{k}+a_{3}\cos^{2}\theta_{k}\right], \nonumber \\
B(\theta_{k},\phi_{k}) &=& \left[b_{1}(a^{2}_{1}-a^{2}_{2})+b_{3}a_{1}a_{3}\right]\sin^{2}\theta_{k} +2a_{3}(b_{1}a_{1}+b_{2}a_{2})\cos^{2}\theta_{k} \nonumber \\
 C &=& a_{3}(a^{2}_{1}-a^{2}_{2}) \eeq with \beq
k^{2} &=& k^{2}_{x}+k^{2}_{y}+k^{2}_{z}, \nonumber \\
\theta_{k} &=& \tan^{-1}(\sqrt{k^{2}_{x}+k^{2}_{y}}/k_{z}), \nonumber \\
\phi_{k} &=& \tan^{-1}(k_{y}/k_{x}). \eeq

Equation (\ref{biquadratic}) is a biquadratic equation with the
following four roots of $k_\ell$ (where $\ell=1,2,3$, or 4) for the
radial wave vectors: \beq\label{biquadratic1}
k^{2}_{1,3} &=& \frac{B+\sqrt{B^{2}-4AC}}{2A}, \nonumber \\
k^{2}_{2,4} &=& \frac{B-\sqrt{B^{2}-4AC}}{2A}. \eeq

So the corresponding $\BM{E}$-field eigenvectors can be obtained
from Eq.~(\ref{matrix}) and are given as follows \beq \BM{E}_{q} &=&
\BM{F}^{e}_{q}f_{q}(\theta_{k},\phi_{k}) = \bigg[
F^{e}_{qx}(\theta_{k},\phi_{k}) \uv{x} +F^{e}_{qy}
(\theta_{k},\phi_{k}) \uv{y} + F^{e}_{qz} (\theta_{k},\phi_{k})
\uv{z} \bigg] f_{q}(\theta_{k},\phi_{k}), \eeq where and
subsequently, $q=1,2,3$, or 4; and\beq\label{eigenvector} F_{qx}^{e}
&=&
-\displaystyle\frac{\triangle_{1}}{\triangle}sin\phi_{k}+\displaystyle\frac{\triangle_{2}}{\triangle}cos\phi_{k},
\nonumber\\
F_{qy}^{e} &=&
\displaystyle\frac{\triangle_{1}}{\triangle}cos\phi_{k}+\frac{\triangle_{2}}{\triangle}sin\phi_{k},\nonumber\\
F_{qz}^{e} &=& 1  \eeq with
 \beq\label{triangle}
 \triangle_{1} &=& i(b_{1}a_{2}+b_{2}a_{1})k^{2}_{q}sin\theta_{k}
cos\theta_{k} \nonumber\\
\triangle_{2} &=&
\left[b_{1}b_{3}k^{2}_{q}sin^{2}\theta_{k}+(b^{2}_{1}-b^{2}_{2})k^{2}_{q}cos^{2}\theta_{k}\right]k^{2}_{q}sin\theta_{k}
cos\theta_{k}-(b_{1}a_{1}+b_{2}a_{2})\nonumber\\
\triangle &=&
-(b_{2}k^{2}_{q}cos^{2}\theta_{k}+a_{2})^{2}+(b_{1}k^{2}_{q}cos^{2}\theta_{k}-a_{1})(b_{1}k^{2}_{q}cos^{2}\theta_{k}+b_{3}k^{2}_{q}sin^{2}\theta_{k}-a_{1})
\eeq With those obtained eigenvalues and their associated formulas,
the $\BM{E}$-field in Eq.~(\ref{fourier}) is then given as follows
\beq\label{E_field} \BM{E}(\BM{r}) &=& \sum^{2}_{q=1} \int^{\pi}_{0}
\int^{2\pi}_{0} \BM{F}^{e}_{q}(\theta_{k},\phi_{k})
f_{q}(\theta_{k},\phi_{k}) e^{i \BM{k}_{q} \cdot \BM{r}}k^{2}_{q}
\sin\theta_{k} d\theta_{k} d\phi_{k} \eeq where \beqq \BM{k}_{q} =
k_{q} \sin\theta_{k} \cos\phi_{k} \uv{x} +k_{q} \sin\theta_{k}
\sin\phi_{k} \uv{y} +k_{q} \cos\theta_{k} \uv{z}, \eeqq and
$f_{q}(\theta_{k},\phi_{k})$ denotes the unknown angular spectrum
amplitude. Equation~(\ref{E_field}) is also known as the eigen plane
wave spectrum representation of the electric field in homogeneous
gyrotropic anisotropic medium. From (\ref{fourier}), it is evident
that the integration over the radial wave-vector component is
reduced to a summation of four terms corresponding to the roots of
(\ref{biquadratic}), which are the only permissible solutions. The
symmetric roots, i.e., $k=-k_{q}$ of $k=k_{q}$ $(q=1, 2)$ are taken
into account automatically as $\theta$ spans from 0 to $\pi$ while
$\phi$ spans from 0 to $2\pi$. Physically, we need to sum up for
only two of the four components, namely, $k_{1}$ and $k_{2}$.

It is noted that the unknown angular spectrum amplitude
$f_{q}(\theta_{k}, \phi_{k})$ is a periodic function with respect to
$\theta_{k}$ and $\phi_{k}$. Therefore we can use surface harmonics
of the first kind to expand the $f_{q}(\theta_{k}, \phi_{k})$
\begin{equation}\label{express}
f_{q}(\theta_{k},\phi_{k}) = \sum_{m',n'} G_{m'n'q} P^{m'}_{n'}
(\cos\theta_{k}) e^{im'\phi_{k}}
\end{equation}
where $P_{n}^{m}(x)$ denotes the associated Legendre function, $n'$
is summed from 0 to $+\infty$, and $m'$ is summed from $-n'$ to
$n'$. Substituting (\ref{express}) to (\ref{E_field}), we obtain
\beq\label{E express} \BM{E}(\BM{r}) \mt&=&\mt \sum^{2}_{q=1}
\sum_{m',n'} G_{m'n'q} \int^{\pi}_{0} \int^{2\pi}_{0} \BM{F}^{e}_{q}
(\theta_{k}, \phi_{k}) P^{m'}_{n'} (\cos\theta_{k}) e^{im'\phi} e^{i
\BM{k}_{q} \cdot \BM{r}} k^{2}_{q} \sin\theta_{k} d\theta_{k}
d\phi_{k}. \eeq This specific form of (\ref{E express}) suggests the
use of the well-known
identity \cite{ref15,ref16}
\beq
e^{i\BM{k} \cdot \BM{r}} \mt&=&\mt \sum^{\infty}_{n=0} i^{n}(2n+1) j_{n}(kr) \bigg[ \sum^{n}_{m=0} \frac{(n-m)!} {(n+m)!} P^{m}_{n} (\cos\theta_{k})P^{m}_{n} (\cos\theta) e^{im(\phi-\phi_{k})}\nonumber \\
&&  +\sum^{n}_{m=1} \frac{(n-m)!} {(n+m)!} P^{m}_{n}
(\cos\theta_{k}) P^{m}_{n}(\cos\theta) e^{-im(\phi-\phi_{k})}
\bigg]. \label{E-series} \eeq After substituting (\ref{E-series})
into (\ref{E express}), we obtain the solution of $\BM{E}(\BM{r})$
for homogeneous gyrotropic anisotropic media. In order to have a
compact and explicit solution to the scattering of a gyrotropic
anisotropic sphere, it is necessary to introduce the spherical
vector wave functions as follows
\cite{ref15} 
\beq\label{Vector function_M} \BM{M}^{(l)}_{mn} &=& z^{(l)}_{n}(kr)
\left [ im \frac{P^{m}_{n} (\cos\theta)}{\sin\theta} \uv{\theta}-
\frac{dP^{m}_{n}(\cos\theta)}{d\theta} \uv{\phi}\right] e^{im\phi} \nonumber \\
\BM{N}^{(l)}_{mn} &=& n(n+1)\frac{z_{n}^{(l)}(kr)}{kr} P^{m}_{n}(\cos\theta) e^{im\phi} \uv{r} + \nonumber\\
&& \frac{1}{kr}\frac{d(rz^{(l)}_{n}(kr))} {dr} \left[ \frac{dP^{m}_{n} (\cos\theta)} {d\theta} \uv{\theta} +im \frac{P^{m}_{n}(\cos\theta)} {\sin\theta} \uv{\phi} \right] e^{im\phi} \nonumber \\
\BM{L}_{mn}^{(l)} &=& k \bigg\{ \frac{dz_{n}^{(l)}(kr)} {d(kr)} P^{m}_{n} (\cos\theta) e^{im\phi} \uv{r}+ \frac{z^{(l)}_{n}(kr)} {kr} \nonumber \\
&&\times \left[ \frac{dP^{m}_{n} (\cos\theta)} {d\theta} \uv{\theta}
+ im \frac{P^{m}_{n} (\cos\theta)} {\sin\theta} \uv{\phi} \right]
e^{im\phi} \bigg\} \eeq where $z^{(l)}_{n}(x)$ (where $l=1$, 2, 3,
or 4) denotes an appropriate kind of spherical Bessel functions,
that is, $j_{n}$, $y_{n}$, $h^{(1)}_{n}$, or $h^{(2)}_{n}$,
respectively. Because of the complete property of the vector wave
functions given in Eq.~(\ref{Vector function_M}), we have the
following expression \beq
\BM{F}^{e}_{q}(\theta,\phi)e^{i\BM{k}_{q}\cdot \BM{r}}\mt&=&\mt\sum_{m,n} \bigg[ A^{e}_{mnq} (\theta_{k}) \BM{M}^{(1)}_{mn} (\BM{r}, k_{q})+ B^{e}_{mnq} (\theta_{k}) \BM{N}^{(1)}_{mn} (\BM{r}, k_{q}) \nonumber \\
&& + C^{e}_{mnq} (\theta_{k}) \BM{L}^{(1)}_{mn} (\BM{r}, k_{q})
\bigg] e^{-im\phi_{k}} \label{result of factor} \eeq where $n$ is
summed from 0 to $+\infty$ while $m$ is summed from $-n$ to $n$, and
$\BM{k}$ is pointing in the ($\theta_{k}, \phi_{k}$) direction while
$\BM{r}$ is pointing in the ($\theta, \phi$) direction in the
spherical coordinates. The other inter-parameters,
$A^{e}_{mnq}(\theta_{k})$, $B^{e}_{mnq}(\theta_{k})$ and
$C^{e}_{mnq}(\theta_{k})$, are provided in Appendix A.

Substituting (\ref{result of factor}) into (\ref{E express}), and
integrating with respect to $\phi_{k}$, we end up with \beq\label{E
end} \BM{E}(\BM{r}) &=& \sum_{q=1}^{2} \sum_{m,n} \sum_{n'} 2\pi
G_{mn'q} \int^{\pi}_{0} \Big[ A^{e}_{mnq}(\theta_{k})
\BM{M}^{(1)}_{mn} (\BM{r}, k_{q}) +B^{e}_{mnq} (\theta_{k})
\BM{N}^{(1)}_{mn} (\BM{r}, k_{q}) \nonumber \\ && +C^{e}_{mnq}
(\theta_{k}) \BM{L}^{(1)}_{mn} (\BM{r}, k_{q}) \Big] P^{m}_{n'}
(\cos\theta_{k})k^{2}_{q} \sin\theta_{k}d\theta_{k}. \eeq
Equation~(\ref{E end}) is the eigenfunction representation of the
$\BM{E}$-field in gyrotropic anisotropic media. The $\BM{H}$-field
eigenvectors can be derived from $\BM{E}$-field eigenvectors shown
in Eqs.~(\ref{biquadratic})-(\ref{biquadratic1}) by using the
source-free Maxwell's equations in the spectral domain. Because the
equations of $\BM{H}$-field are very similar to those of
$\BM{E}$-field, we only give the relation between $\BM{H}$-field
eigenvectors (i.e., $\BM{F}_{q}^{h}$) and $\BM{E}$-field
eigenvectors (i.e., $\BM{F}_{q}^{e}$) in Cartesian coordinates \beq
\BM{F}_{q}^{h} \mt&=&\mt\frac{k_{q}}{\omega}\left[
\begin{array}{ccc}
b_{1} & ib_{2} & 0 \\ -ib_{2} & b_{1} & 0 \\
0 & 0 & b_{3}
\end{array}\right] \left[
\begin{array}{ccc}
0 & -\cos\theta_{k} & \sin\theta_{k} \sin\phi_{k}\\ \cos\theta_{k} & 0 & -\sin\theta_{k} \cos\phi_{k} \\
-\sin\theta_{k} \sin\phi_{k} & \sin\theta_{k} \cos\phi_{k} & 0
\end{array}\right]\cdot  \BM{F}_{q}^{e} \nonumber\\ \eeq where $q=1,2$. 

From the result shown in (\ref{E end}), it can be seen that the
solutions to the source-free Maxwell's equations for the gyrotropic
anisotropic medium are expanded in terms of the first kind of
spherical vector functions. Because all spherical Bessel functions
of different kinds satisfy the same differential equation and the
same recursive relations, we can use the field expressions given in
Eq.~(\ref{E end}) to analyze scattering and radiation by the stacked
structure of the gyrotropic anisotropic media.

Assume that the electric field of an incident plane wave is given by
$\BM{E}=\uv{x}E_{0} e^{ik_{0}z}$. The incident EM fields (designated
by the superscript $inc$) can be expanded by an infinite series of
spherical vector wave functions for an isotropic medium as follows
\cite{ref13}: \beq
\BM{E}^{inc} &=& E_{0}\sum_{m,n} [\delta_{m,1}+\delta_{m,-1}] \Big[ a^{x}_{mn} \BM{M}_{mn}^{(1)} (\BM{r}, k_{0})+ b^{x}_{mn} \BM{N}_{mn}^{(1)} (\BM{r}, k_{0})\Big] \nonumber \\
\BM{H}^{inc} &=& \frac{k_{0}}{i\omega \mu_{0}} E_{0} \sum_{mn}
[\delta_{m,1}+\delta_{m,-1}] \Big[a^{x}_{mn}{\bf N}_{mn}^{(1)}
(\BM{r}, k_{0})+b^{x}_{mn} \BM{M}_{mn}^{(1)}(\BM{r}, k_{0}) \Big]
\eeq where \beq
a^{x}_{mn} &=& \left\{\begin{array}{ll} \displaystyle i^{n+1}\frac{2n+1}{2n(n+1)}, & m=1 \\ \displaystyle i^{n+1}\frac{2n+1}{2}, & m=-1; \end{array}\right. \nonumber \\
b^{x}_{mn} &=&\left\{ \begin{array}{ll} \displaystyle i^{n+1}\frac{2n+1}{2n(n+1)}, & m=1 \\ \displaystyle -i^{n+1}\frac{2n+1}{2}, & m=-1 \end{array}\right. \nonumber\\
\delta_{s,l} &=&\left\{\begin{array}{ll} 1 & s=l \\ 0 & s\neq l
\end{array}\right. . \eeq According to the radiation condition
of an outgoing wave and asymptotic behavior of spherical Bessel
functions, only $h_{n}^{(1)}$ should be retained in the radial
function, therefore the scattering fields (designated by the
superscript $s$) are expanded as \beq
\BM{E}^{s}&=&\sum_{mn}\left[A^{s}_{mn}\BM{M}_{mn}^{(3)}(\BM{r},
k_{0})
+B^{s}_{mn} \BM{N}_{mn}^{(3)} (\BM{r},k_{0})\right]\nonumber\\
\BM{H}^{s}&=&\frac{k_{0}}{i\omega \mu_{0}} \sum_{mn} \left[
A^{s}_{mn} \BM{N}_{mn}^{(3)}(\BM{r}, k_{0}) +B^{s}_{mn}
\BM{M}_{mn}^{(3)} (\BM{r}, k_{0})\right] \eeq where $A^{s}_{mn}$ and
$B^{s}_{mn}$ (with $n$ being from 0 to $+\infty$ and $m$ being from
$-n$ to $n$) are unknown coefficients, and $k_{0} =\omega(
\epsilon_{0} \mu_{0} )^{1/2}$, $\epsilon_{0}$ and $\mu_{0}$ denote
the wave number, permittivity and permeability in free space,
respectively.

The expressions of EM fields inside the gyrotropic anisotropic
sphere are given in Eq.~(\ref{E end}), and the continuity of the
tangential EM field components at $r=a$ yields \beq\label{continue1}
&& \sum_{q=1}^{2}\sum_{n'=0}^{\infty}2\pi G_{mn'q} \int_{0}^{\pi} Q_{mnq} P^{m}_{n'} (\cos\theta_{k})k^{2}_{q} \sin\theta_{k} d\theta_{k} \nonumber\\
&&=E_{0} \left[ \delta_{m,1} +\delta_{m,-1} \right] a^{x}_{mn} \cdot \frac{i}{(k_{0}a)^{2}} \nonumber \\
&& \sum_{q=1}^{2}\sum_{n'=0}^{\infty}2\pi G_{mn'q} \int_{0}^{\pi}
R_{mnq} P^{m}_{n'} (\cos\theta_{k}) k^{2}_{q} \sin\theta_{k}
d\theta_{k}\nonumber\\
&&=E_{0} \left[ \delta_{m,1} +\delta_{m,-1} \right] b^{x}_{mn} \cdot
\frac{i}{(k_{0}a)^{2}}  \eeq where \beq\label{continue1_factor}
Q_{mnq} &=& \bigg\{ A_{mnq}^{e} \frac{1}{k_{0}r} \frac{d}{dr} \left[ rh^{(1)}_{n}(k_{0}r) \right] j_{n}(k_{q}r) \nonumber \\
&& - \frac{i\omega \mu_{0}} {k_{0}} \bigg[ B^{h}_{mnq} \frac{1}{k_{q}r} \frac{d}{dr} \left[ rj_{n} (k_{q}r) \right] \nonumber \\
&& + C^{h}_{mnq} \frac{j_{n} (k_{q}r)} {r} \bigg] \cdot h^{(1)}_{n} (k_{0}r) \bigg\}_{r=a} \nonumber \\
R_{mnq} &=&\bigg\{ \frac{i\omega \mu_{0}} {k_{0}} A_{mnq}^{h} \frac{1}{k_{0}r} \frac{d}{dr} \left(rh^{(1)}_{n} (k_{0}r) \right) j_{n}(k_{q}r) \nonumber \\
&& - \bigg[ B^{e}_{mnq} \frac{1}{k_{q}r} \frac{d}{dr} \left(rj_{n}(k_{q}r) \right) \nonumber \\
&& +C^{e}_{mnq} \frac{j_{n}(k_{q}r)}{r} \bigg] \cdot
h^{(1)}_{n}(k_{0}r) \bigg\}_{r=a}.   \eeq The scattering
coefficients, i.e., $A^{s}_{mn}$ and $B^{s}_{mn}$, are thus
expressed as \beq
A^{s}_{mn} &=& \frac{1}{h^{(1)}_{n}(k_{0}a)} \bigg[ \sum^{\infty}_{n'=0} \sum^{2}_{q=1} 2\pi G_{mnq} \int^{\pi}_{0} A_{mnq}^{e} j_{n}(k_{q}a) P_{n'}^{m} k^{2}_{q} \sin\theta_{k} d\theta_{k}\nonumber\\
&& -E_{0}[ \delta_{m,1} +\delta_{m,1}] a^{x}_{mn}j_{n}(k_{0}a) \bigg]\nonumber \\
B^{s}_{mn} &=& \frac{1}{h^{(1)}_{n}(k_{0}a)} \bigg[ \frac{i\omega \mu_{0}}{k_{0}} \sum^{\infty}_{n'=0} \sum^{2}_{q=1} 2\pi G_{mnq} \int^{\pi}_{0} A_{mnq}^{h} j_{n}(k_{q}a) P_{n'}^{m} k^{2}_{q} \sin\theta_{k} d\theta_{k} \nonumber \\
&& -E_{0}[\delta_{m,1}+\delta_{m,1}] b^{x}_{mn} j_{n}(k_{0}a)
\bigg]. \eeq From those determined scattering coefficients, the
radar cross sections (RCSs) of the gyrotropic anisotropic sphere can
be calculated, i.e., \beq
\sigma &=& \lim_{r\longrightarrow\infty} 4\pi r^{2}\frac{|E^{s}|^{2}}{|E^{i}|^{2}} \nonumber \\
&=& \frac{4\pi}{E^{2}_{0}k_0^{2}} \Bigg[ \bigg| \sum_{n=1}^{\infty} (-i)^{n} \bigg\{ \frac{P^{1}_{n}}{\sin\theta} \left[ A^{s}_{1n} e^{i\phi} + \frac{A^{s}_{-1n}}{n(n+1)} e^{-i\phi} \right] \nonumber \\
&& +\frac{dP^{1}_{n}}{d\theta} \left[ B^{s}_{1n} e^{i\phi} - \frac{B^{s}_{-1n}}{n(n+1)} e^{-i\phi} \right] \bigg\} \bigg|^{2} \nonumber \\
&& + \Big|\sum_{n=1}^{\infty}(-i)^{n+1} \bigg\{ \frac{dP^{1}_{n}}{d\theta} \left[ A^{s}_{1n} e^{i\phi} - \frac{A^{s}_{-1n}}{n(n+1)} e^{-i\phi}\right] \nonumber \\
&& + \frac{P^{1}_{n}}{\sin\theta} \left[ B^{s}_{1n} e^{i\phi} +
\frac{B^{s}_{-1n}}{n(n+1)} e^{-i\phi}\right] \bigg\} \bigg|^{2}
\Bigg]. \eeq

\section{Numerical Results and Discussion}

\begin{figure}[htbh]
\centering\includegraphics[width=6.5cm]{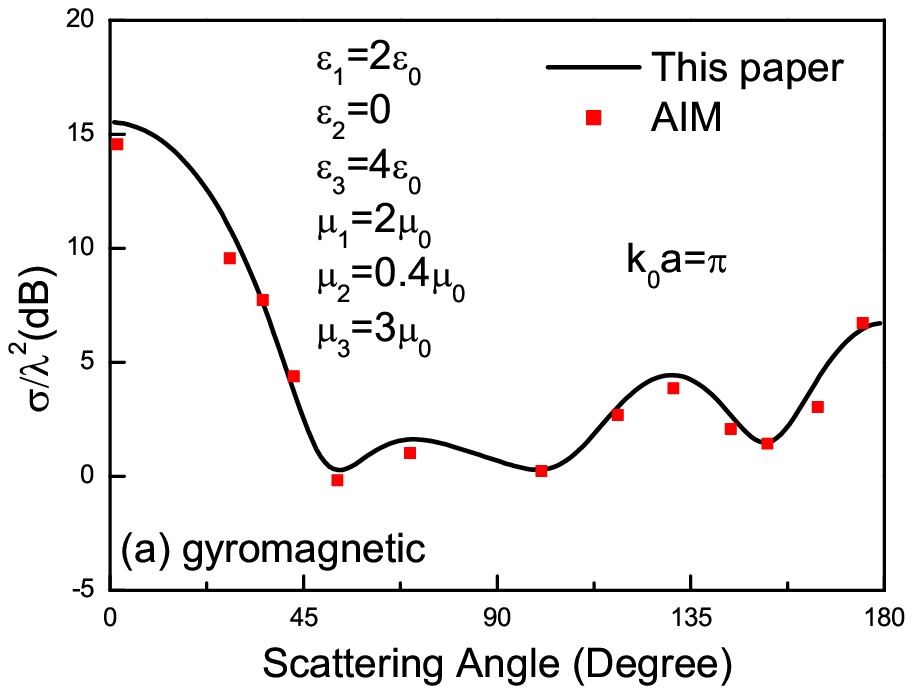}\\
\centering\includegraphics[width=6.5cm]{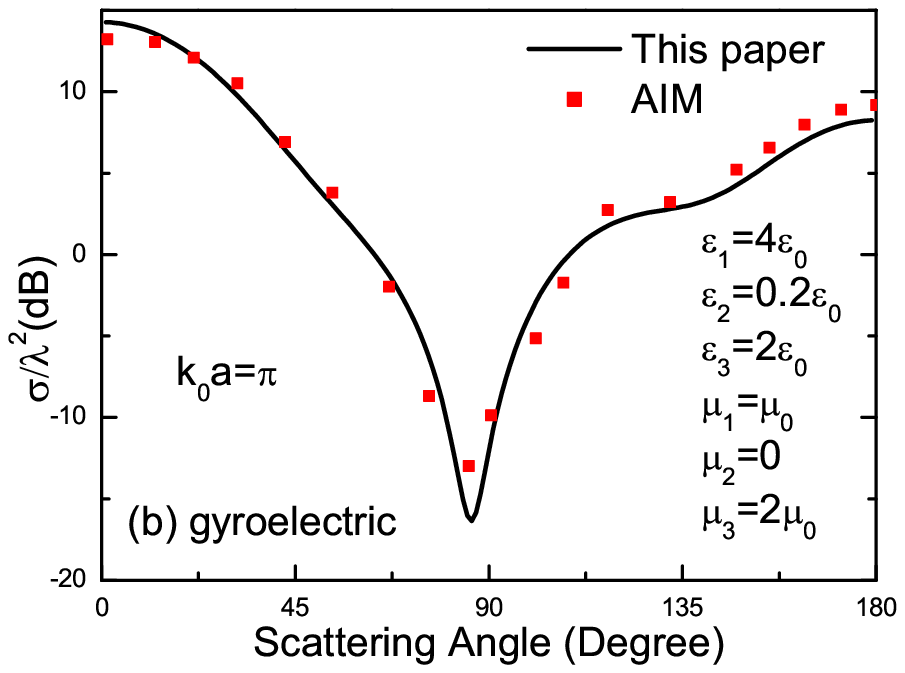}  \caption{Radar
cross sections (RCSs) versus the scattering angle
 (in
degree) for (a) the gyromagnetic sphere and (b) the gyroelectric
sphere. The comparisons in RCS results are made between our
spectral-domain method (solid curve) and the AIM (square dot). The
electronic size is fixed at $k_{0}a=\pi$.} \label{fig2}
\end{figure}

To verify this spectral-domain scattering method for the gyrotropic
anisotropic sphere, we present the bistatic radar cross sections
(RCSs) in $E$-plane ($xoz$-plane as shown in Fig.~\ref{fig1}) and
$H$-plane ($yoz$-plane as shown in Fig.~\ref{fig1}) which are
compared to the results calculated by a numerical algorithm, i.e.,
adaptive integral method (AIM) \cite{ref16b} extended from Ref[17].
The gyromagnetic ($\epsilon_2=0$ and $\mu_2\neq 0$ in
Fig.~\ref{fig2}(a)) and gyroelectric ($\epsilon_2\neq 0$ and
$\mu_2=0$ in Fig.~\ref{fig2}(a)) cases have been discussed in
Fig.~\ref{fig2}, and the good agreement of RCS results on both
planes is achieved between our method and AIM. It partially verifies
that the proposed method and the Fortran code developed in this
paper are correct. The series in (\ref{continue1}) converge rapidly,
and it is sufficient to take $N=4$ as the upper limit of the
summation indices $n$ and $n'$. Certainly, it should be pointed out
that the convergence rate or the upper limit of the summation
depends on the electrical dimension of the sphere (with respect to
the wavelength).

\begin{figure}[htbh]
\centering\includegraphics[width=7.5cm]{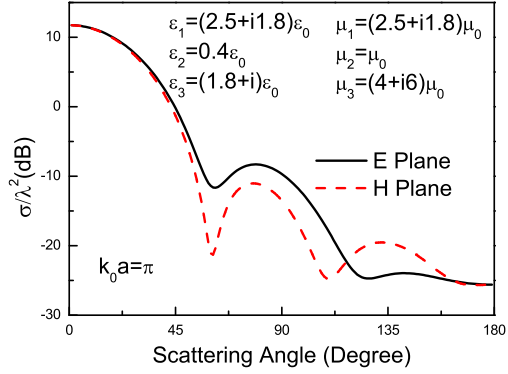} \caption{Radar
cross sections (RCSs) versus scattering angle
 (in
degrees): The electronic size is chosen as $k_{0}a=\pi$.}
\label{fig3}
\end{figure}

Then we study a more general case in Fig.~\ref{fig3} in which both
material tensors ($\dy{\epsilon}$ and $\dy{\mu}$) are gyrotropic and
lossy. The radar cross sections on $E$-plane and $H$-plane have been
shown in Fig.~\ref{fig3}. To the best of our knowledge, the
scattering by such a general gyrotropic sphere has not been
reported, except for its subcase of gyromagnetic spheres
\cite{ref12b}. Obviously, our model is more general in terms of the
material complexity in Ref[13]. Our spectral-domain analysis is
distinguished from the spatial-domain method in \cite{ref12b}, and
one can imagine that if the spatial method in Ref[13] is extended to
study our general gyrotropic materials, the formulation would be
lengthy due to the second tensor of permittivity. Hence, even for
the general gyrotropic materials, our method results in simplified
formulation.

\begin{figure}[htbh]
\centering\includegraphics[width=7.5cm]{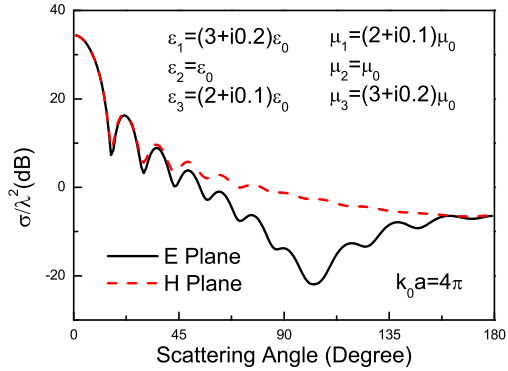} \caption{Radar
cross sections (RCSs) versus scattering angle $\theta$ (in degree)
in $E$-plane (solid curve) and and $H$-plane (dashed curve).
 The electric dimension is chosen to be $k_0a =4\pi$..} \label{fig4}
\end{figure}

To illustrate the applicability of this analytical solution to the
gyrotropic anisotropic sphere of electrically large size (for
example, in the resonance region), the RCSs of a relatively large
gyrotropic anisotropic sphere with loss are presented in Fig.~4. The
lossy permittivity and permeability parameters are chosen as
$\epsilon_{1} =(3+0.2i)\epsilon_0$, $\epsilon_{2} =\epsilon_0$,
$\epsilon_{3} =(2+0.1i)\epsilon_0$, $\mu_{1} =(2+0.1i)\mu_0$,
$\mu_{2} =\mu_0$, $\mu_{3} =(3+0.2i)\mu_0$. When the dimensions are
increased, the convergence number ($N=24$ for the sphere
$k_{0}a=4\pi$ in Fig.~4) is also increased.

\section{Conclusions}

In this paper, an analytical solution to the scattering by a general
gyrotropic anisotropic sphere has been obtained. The method is
developed based on the multipole expansion of the field along with
the Fourier transform where the unknown angular spectrum amplitude
is determined in spectral domain. The three-dimensional
electromagnetic scattering of a plane wave by an gyrotropic
anisotropic sphere has been theoretically formulated, physically
characterized and numerically discussed. Numerical results for
special cases are also obtained and verified by comparing with the
results from the method of moments. The good agreement validates our
spectral-domain scattering theory. By using our proposed theory, the
scattering problems of the general optically anisotropic sphere can
be analytically studied in spectral domain and RCSs can be readily
computed. The analytical solution under arbitrary incident angle is
still under investigation.


\appendix
\setcounter{equation}{0}
\renewcommand{\theequation}{A-\arabic{equation}}

\section{Scattering coefficients of eigen-expansions in Eqs.~(20) and (27)}

\beq
{\bf \BM{F}}^{e}_{q}(\theta,\phi) e^{i{\bf \BM{k}}_{q}\cdot {\bf \BM{r}}} \mt&=&\mt \sum_{mn} \Big[ A^{e}_{mnq}(\theta_{k}) {\bf \BM{M}}^{(1)}_{mn} ({\bf \BM{r}},k_{q}) \nonumber \\
&&\mt + B^{e}_{mnq} (\theta_{k}) {\bf \BM{N}}^{(1)}_{mn} ({\bf \BM{r}},k_{q}) \nonumber \\
&&\mt + C^{e}_{mnq}(\theta_{k}) {\bf \BM{L}}^{(1)}_{mn} ({\bf
\BM{r}},k_{q}) \Big] e^{-im\phi_{k}}. \quad\quad \eeq

Because the spherical wave functions ${\bf \BM{L}}_{mn} ({\bf
\BM{r}},k)$, ${\bf \BM{M}}_{mn} ({\bf \BM{r}},k)$, and ${\bf
\BM{N}}_{mn} ({\bf \BM{r}},k)$ form a complete set of orthogonal
basis functions, we can employ them to expand any solutions
uniquely, {\em e.g.}, \beq\label{plane3}
\widehat{x} e^{i{\bf \BM{k}}_{q} \cdot {\bf \BM{r}}} \mt \ \ &=& \ \ \mt \sum_{mn} \Big[ a^{x}_{mn} (\theta_{k}) {\bf \BM{M}}^{(1)}_{mn} ({\bf \BM{r}},k_{q}) + b^{x}_{mn} (\theta_{k}) \quad\quad \nonumber \\
&&\mT \cdot {\bf \BM{N}}^{(1)}_{mn} ({\bf \BM{r}},k_{q}) + c^{x}_{mn} (\theta_{k}) {\bf \BM{L}}^{(1)}_{mn} ({\bf \BM{r}},k_{q}) \Big], \nonumber \\
\widehat{y} e^{i{\bf \BM{k}}_{q} \cdot {\bf \BM{r}}} \mt \ \ &=& \ \ \mt \sum_{mn} \Big[ a^{y}_{mn} (\theta_{k}) {\bf \BM{M}}^{(1)}_{mn} ({\bf \BM{r}},k_{q}) + b^{y}_{mn} (\theta_{k}) \nonumber \\
&&\mT \cdot {\bf \BM{N}}^{(1)}_{mn} ({\bf \BM{r}},k_{q}) + c^{y}_{mn} (\theta_{k}) {\bf \BM{L}}^{(1)}_{mn}({\bf \BM{r}},k_{q}) \Big], \nonumber  \\
\widehat{z}e^{i{\bf \BM{k}}_{q}\cdot {\bf \BM{r}}} \mt \ \ &=& \ \ \mt \sum_{mn} \Big[ a^{z}_{mn} (\theta_{k}) {\bf \BM{M}}^{(1)}_{mn} ({\bf \BM{r}},k_{q}) + b^{z}_{mn} (\theta_{k}) \nonumber \\
&&\mT \cdot {\bf \BM{N}}^{(1)}_{mn} ({\bf \BM{r}},k_{q}) +
c^{z}_{mn} (\theta_{k}) {\bf \BM{L}}^{(1)}_{mn} ({\bf \BM{r}},k_{q})
\Big].  \eeq

The coefficients in (\ref{plane3}), i.e., $a^{p}_{mn}$, $b^{p}_{mn}$
and $c^{p}_{mn}$ (where $p=x,y,z$), are functions of $\theta_{k}$
and $\phi_{k}$. For the detailed expansion and discussion, the
information can be found in \cite{ref13}. We provide only the
coefficients of $A^{e}_{mnq}$, $B^{e}_{mnq}$ and $C^{e}_{mnq}$ used
in the main text. From Eq.~(\ref{eigenvector}), we have \be {\bf
\BM{F}}^{e}_{q}(\theta_{k},\phi_{k})={\bf
\BM{F}}^{e1}_{q}(\theta_{k},\phi_{k})+{\bf
\BM{F}}^{e2}_{q}(\theta_{k},\phi_{k}), \ee where \beq {\bf
\BM{F}}^{ep}_{q}(\theta_{k},\phi_{k}) =
F^{ep}_{qx}(\theta_{k},\phi_{k})\widehat{x}
+F^{ep}_{qy}(\theta_{k},\phi_{k})\widehat{y} +
F^{ep}_{qz}(\theta_{k},\phi_{k})\widehat{z}~~~~~~(p=1,~2) \eeq with
\beq
F^{ep}_{qx}(\theta_{k},\phi_{k}) \mt \ \ &=& \ \ \mt \left\{ \begin{array}{ll} \displaystyle -\frac{\triangle_{1}}{\triangle} \sin\phi_{k}, & p=1, \\
\displaystyle \frac{\triangle_{2}}{\triangle} \cos\phi_{k}, & p=2; \end{array}\right. \nonumber \\
F^{ep}_{qy}(\theta_{k},\phi_{k}) \mt \ \ &=& \ \ \mt \left\{ \begin{array}{ll} \displaystyle \frac{\triangle_{1}}{\triangle} \cos\phi_{k}, & p=1, \\
\displaystyle \frac{\triangle_{2}}{\triangle} \sin\phi_{k}, & p=2; \end{array}\right.\nonumber  \\
F^{ep}_{qz}(\theta_{k},\phi_{k}) \mt \ \ &=& \ \ \mt \left\{
\begin{array}{ll} 0, & p=1, \\ 1, & p=2. \end{array}\right. \eeq

In the above equations, the intermediate parameters,
$\triangle_{1}$, $\triangle_{2}$ and $\triangle$, are functions of
only $\theta_{k}$ as given in Eq.~(\ref{triangle}). Then we can
split the parameters as follows \beq
A^{e}_{mnq} \mt \ \ &=& \ \ \mt A^{e1}_{mnq}+A^{e2}_{mnq},\nonumber \\
B^{e}_{mnq} \mt \ \ &=& \ \ \mt B^{e1}_{mnq}+B^{e2}_{mnq},\nonumber  \\
C^{e}_{mnq} \mt \ \ &=& \ \ \mt C^{e1}_{mnq}+C^{e2}_{mnq}, \eeq and
thus obtain ($p=1$ or 2) \beq
A^{ep}_{mnq}e^{-im\phi_{k}} \mt \ \ &=& \ \ \mt F^{ep}_{qx}a^{x}_{mn}+F^{ep}_{qy}a^{y}_{mn}+F^{ep}_{qz}a^{z}_{mn}, \quad\quad \nonumber \\
B^{ep}_{mnq}e^{-im\phi_{k}} \mt \ \ &=& \ \ \mt F^{ep}_{qx}b^{x}_{mn}+F^{ep}_{qy}b^{y}_{mn}+F^{ep}_{qz}b^{z}_{mn}, \nonumber \\
C^{ep}_{mnq}e^{-im\phi_{k}} \mt \ \ &=& \ \ \mt
F^{ep}_{qx}c^{x}_{mn}+F^{ep}_{qy}c^{y}_{mn}+F^{ep}_{qz}c^{z}_{mn}.
\eeq
As a result, we can now obtain the expansion coefficients of ${\bf E}$-fields in a gyrotropic anisotropic medium, {\em i.e.}, $A^{ep}_{mnq}$, $B^{ep}_{mnq}$ and $C^{ep}_{mnq}$ (where $q=1,2$), as follows: \\
for $p=1$ and $m\geq 0$ \beq
A^{e1}_{mnq} \mt \ \ &=& \ \ \mt i^{n} \frac{2n+1}{2n(n+1)} \frac{(n-m)!}{(n+m)!} \frac{\triangle_{1}}{\triangle} \Big[ (n+m) (n-m \nonumber \\
&&\mt +1) P^{m-1}_{n} (\cos\theta_{k}) -P^{m+1}_{n} (\cos\theta_{k}) \Big], \nonumber \\
B^{e1}_{mnq} \mt \ \ &=& \ \ \mt i^{n} \frac{1}{2n(n+1)} \frac{(n-m)!}{(n+m)!} \frac{\triangle_{1}}{\triangle} \Big[ (n+1) (n+m) \nonumber \\
&&\mt \times (n+m-1) P^{m-1}_{n-1} (\cos\theta_{k}) + (n+1) \nonumber \\
&&\mt \times P^{m+1}_{n-1} (\cos\theta_{k}) + n (n\!-\!m\!+\!2) (n\!-\!m\!+\!1) \nonumber \\
&&\mt \times P^{m-1}_{n+1} (\cos\theta_{k}) + n P^{m+1}_{n+1} (\cos\theta_{k}) \Big], \nonumber \\
C^{e1}_{mnq} \mt \ \ &=& \ \ \mt i^{n} \frac{1}{2k_{q}} \frac{(n-m)!}{(n+m)!} \frac{\triangle_{1}}{\triangle} \Big[ (n+m) (n+m-1) \nonumber \\
&&\mt \times P^{m-1}_{n-1} (\cos\theta_{k}) + P^{m+1}_{n-1} (\cos\theta_{k}) \nonumber \\
&&\mt - (n-m+2) (n-m+1) P^{m-1}_{n+1} (\cos\theta_{k}) \nonumber \\
&&\mt - P^{m+1}_{n+1} (\cos\theta_{k}) \Big]; \eeq while for $p=1$
and $m>0$, \beq
A^{e1}_{-mnq} \mt \ \ &=& \ \ \mt (-1)^{m}\frac{(n+m)!}{(n-m)!} A^{e1}_{mnq}, \nonumber \\
B^{e1}_{-mnq} \mt \ \ &=& \ \ \mt (-1)^{m+1}\frac{(n+m)!}{(n-m)!}B^{e1}_{mnq}, \quad\quad \nonumber \\
C^{e1}_{-mnq} \mt \ \ &=& \ \ \mt
(-1)^{m+1}\frac{(n+m)!}{(n-m)!}C^{e1}_{mnq}. \eeq Similarly, for
$p=2$ and $m\geq 0$, we have \beq
A^{e2}_{mnq} \mt \ \ &=& \ \ \mt i^{n+1}\displaystyle\frac{2n+1}{n(n+1)}\frac{(n-m)!}{(n+m)!}\Big\{ \frac{\triangle_{2}}{2\triangle} \Big[ (n\!+\!m) (n\!-\!m \nonumber \\
&&\mt +1) P^{m-1}_{n} (\cos\theta_{k}) + P^{m+1}_{n} (\cos\theta_{k}) \Big] \nonumber \\
&&\mt + m P^{m}_{n} (\cos\theta_{k}) \Big\}, \nonumber \\
B^{e2}_{mnq} \mt \ \ &=& \ \ \mt i^{n+1}\displaystyle\frac{1}{n(n+1)}\frac{(n-m)!}{(n+m)!} \Big\{ \frac{\triangle_{2}}{2\triangle} \Big[ (n\!+\!1) (n\!+\!m) \nonumber \\
&&\mt \times (n\!+\!m\!-\!1) P^{m-1}_{n-1} (\cos\theta_{k}) - (n+1) \nonumber \\
&&\mt \times P^{m+1}_{n-1} (\cos\theta_{k}) + n (n\!-\!m\!+\!2) (n\!-\!m\!+\!1) \nonumber \\
&&\mt \times P^{m-1}_{n+1} (\cos\theta_{k}) - nP^{m+1}_{n+1} (\cos\theta_{k}) \Big] \nonumber \\
&&\mt + \Big[ n (n-m+1) P^{m}_{n+1} (\cos\theta_{k}) \nonumber \\
&&\mt - (n+1) (n+m) P^{m-1}_{n-1} (\cos\theta_{k}) \Big] \Big\}, \nonumber \\
C^{e2}_{mnq} \mt \ \ &=& \ \ \mt i^{n+1}\displaystyle\frac{1}{k_{q}}\frac{(n-m)!}{(n+m)!} \Big\{ \frac{\triangle_{2}}{2\triangle} \Big[ (n+m) (n+m-1) \nonumber \\
&&\mt \times P^{m-1}_{n-1} (\cos\theta_{k}) - P^{m+1}_{n-1} (\cos\theta_{k}) \nonumber \\
&&\mt - (n-m+2) (n-m+1) P^{m-1}_{n+1} (\cos\theta_{k}) \nonumber \\
&&\mt + P^{m+1}_{n+1} (\cos\theta_{k}) \Big] -(2n+1) \cos\theta_{k} \nonumber \\
&&\mt \times P^{m}_{n} (\cos\theta_{k}) \Big\}; \eeq while for $p=2$
and $m>0$, \beq
A^{e2}_{-mnq} \mt\ \ &=&\ \ \mt (-1)^{m+1}\frac{(n+m)!}{(n-m)!}A^{e2}_{mnq}, \quad\quad \nonumber \\
B^{e2}_{-mnq} \mt\ \ &=&\ \ \mt (-1)^{m}\frac{(n+m)!}{(n-m)!}B^{e2}_{mnq}, \nonumber \\
C^{e2}_{-mnq} \mt\ \ &=&\ \ \mt
(-1)^{m}\frac{(n+m)!}{(n-m)!}C^{e2}_{mnq}. \eeq In a procedure
similar to the above, the expansion coefficients of the ${\bf
H}$-field eigenvector in gyrotropic anisotropic medium can be also
obtained.

\section*{Acknowledgments}
The authors are grateful for the support from National University of
Singapore under the Grant No. R-263-000-574-133. This work is
partially supported by the Grant No. 60971047 of National Natural
Science Foundation of China (NSFC), and the Grant No. Y1080730 of
Natural Science Foundation of Zhejiang Province.

\end{document}